\documentclass[prb,aps,showpacs,floats,twocolumn]{revtex4}

\usepackage[dvips]{color}

\usepackage{amsmath}
\usepackage{graphicx}
\usepackage{amsfonts}
\usepackage{amssymb}

\newcommand{\bqa}{\begin{eqnarray}}
\newcommand{\eqa}{\end{eqnarray}}

\begin{document}

\title{Probing surface states of topological insulator:
Kondo effect and Friedel oscillation under magnetic field}

\author{Minh-Tien Tran$^{1,2}$ and Ki-Seok Kim$^{1,3}$}
\affiliation{ $^1$Asia Pacific Center for Theoretical Physics,
Pohang, Gyeongbuk 790-784,
Republic of Korea \\
$^2$Institute of Physics, Vietnamese Academy of Science and
Technology, P.O.Box 429, 10000 Hanoi, Vietnam \\
$^3$Department of Physics, Pohang University of Science and
Technology, Pohang, Gyeongbuk 790-784, Korea}

\begin{abstract}
We address three issues on the role of magnetic impurities in the
surface state of a three dimensional topological insulator. First,
we prove that the Kondo effect of the topological surface is
essentially the same as that of the graphene surface,
demonstrating that an effective impurity action of the topological
surface coincides with that of the graphene surface. Second, we
study the role of the $z$-directional magnetic field ($h$) in the
Kondo effect, and show that the peak splitting in the impurity
local density of states does not follow the $h$-linear behavior,
the typical physics in the soft-gap Kondo model. We discuss that
the origin is spin locking in the helical surface. Third, we
examine the Friedel oscillation around the magnetic impurity. It
turns out that the pattern of Friedel oscillation in the helical
surface is identical to that of the graphene surface, displaying
the inverse-square behavior $\sim r^{-2}$ if the inter-valley
scattering is not introduced in the graphene case. However,
introduction of the magnetic field leads the electron density of
states from the inverse-square physics to the inverse behavior
$\sim r^{-1}$ in the topological insulator's surface while it
still remains as $\sim r^{-2}$ in the graphene Kondo effect. We
discuss that this originates from spin flipping induced by
magnetic field.

\end{abstract}


\maketitle

\section{Introduction}

It has been one of the main research interests to understand
topological states of matter in modern condensed matter
physics.\cite{Review_Topology} Electrons in two dimensions under
strong magnetic field form such a topological state, exhibiting
the quantum Hall effect (QHE) with gapless chiral edge modes
protected by topology.\cite{Review_QHE} Later, Haldane proposed an
interesting toy model to show the QHE without Landau levels,
basically the tight-binding model for spinless electrons in the
honeycomb lattice with a complex next-nearest-neighbor hopping
parameter.\cite{Haldane_Model} Recently, Kane and Mele extended
the Haldane model into the case of spinful electrons, where each
spin observes an opposite fictitious magnetic flux, realized by
the spin-orbit coupling and preserving time reversal
symmetry.\cite{Kane_Mele} Analogous with the QHE, this insulating
state shows the spin QHE, where the difference of the spin
$\uparrow$ and $\downarrow$ Hall conductances is quantized. When
the spin quantum number is not conserved due to the presence of
the Rashba-type spin-orbit interaction, the spin Hall conductance
cannot be used for topological characterization. Kane and Mele
proposed the Z$_{2}$ index, counting the number of helical edge
states with modular $2$, and concluded that the odd Z$_{2}$ index
state corresponds to a topological state of an insulator, which
cannot be adiabatically connected to the even Z$_{2}$ state of a
trivial insulator.\cite{Kane_Mele}

Immediately, it was performed three dimensional generalization of
the two dimensional Z$_{2}$ topological
insulator.\cite{Fu,Balents,Roy} It turns out that the surface
state of the three dimensional topological insulator has an odd
number of helical Dirac fermions,\cite{Zhang1} identified with an
odd Z$_{2}$ index, where spins are locked along the momentum
direction on the surface. Since there is no QHE analogue in three
dimensions, the three dimensional topological insulator has been
regarded as a new quantum state of matter.

An interesting prediction is associated with the stability of the
helical metallic state against time reversal symmetric
perturbation, where it should be stable against Anderson
localization independent of the disorder strength.\cite{Ryu} In
addition, the topological $\theta$ term appears due to the chiral
anomaly in the field theory context, giving rise to electric
polarization along the magnetic field direction when time reversal
symmetry breaking field is applied
infinitesimally.\cite{Zhang2,Moore}

Although the topological insulator was now verified
experimentally,\cite{Hsieh1,Hsieh2,Hsieh3,Chen} the angle reserved
photo emission spectroscopy (ARPES) is the only experimental
probe, not performed on the table top. In this respect it is
desirable to propose table top experiments for three dimensional
topological insulators. Recently, an experimental realization of
the magnetic doping on the surface of three-dimensional
topological insulator was reported.\cite{exp_kondo}

In this paper we propose how to probe the surface state of the
topological insulator, investigating the Kondo effect and Friedel
oscillation under magnetic field. Since spins of host electrons
are locked along the momentum in the topological surface, one may
expect under-screening for a magnetic impurity in contrast with
the soft-gap Kondo physics \cite{Cassanello} expected to arise in
the graphene surface.\cite{Neto} However, it turns out that the
Kondo effect of the topological surface is essentially the same as
that of the graphene surface. Integrating over host electrons, we
also prove that an effective impurity action of the helical metal
coincides with that of the soft-gap Anderson model. This precise
demonstration tells us exact impurity screening at hybridization
enough to overcome the vanishing density of states. In order to
distinguish the Kondo effect of the topological surface from the
soft-gap Kondo effect, we introduce magnetic field in the
$z$-direction ($h$). We show that the peak splitting in the
impurity local density of states does not follow the $h$-linear
behavior, the typical physics in the soft-gap Anderson or Kondo
model. We derive an analytic expression based on the picture of
spin locking in the helical surface.

Finally, we examine the feedback effect of the impurity dynamics
to host electrons, evaluating the host-electron local density of
states (LDOS) and revealing the Friedel oscillation around the
magnetic impurity. The LDOS is measurable by Fourier
transformation scanning tunneling spectroscopy (FTSTS). The FTSTS
measurements have been developed for graphene,\cite{Mallet,Rutter}
as well as other two dimensional materials.\cite{Vonau,Dupont} The
LDOS in graphene was also investigated theoretically.\cite{Bena}
It turns out that the pattern of Friedel oscillation in the
topological surface is identical to that of the graphene surface
without magnetic field, displaying the inverse-square behavior
$\sim r^{-2}$ if the inter-valley scattering is not introduced in
the graphene case, where $r$ is the distance from the impurity
position. However, introduction of the magnetic field leads the
LDOS from the inverse-square physics to the inverse behavior $\sim
r^{-1}$ in the topological insulator's surface while it still
remains as $\sim r^{-2}$ in the graphene Kondo effect. We discuss
that this originates from spin flipping induced by magnetic field.

This study addresses three issues on the role of magnetic
impurities in the surface state of a three dimensional topological
insulator. First, we prove that the Kondo effect of the
topological surface is essentially the same as that of the
graphene surface. Second, we study the role of the $z$-directional
magnetic field ($h$) in the Kondo effect, which turns out to
differ from the typical physics in the soft-gap Kondo model.
Third, we examine the effect of the magnetic field on the Friedel
oscillation around the magnetic impurity. The previous studies of
Refs. \onlinecite{Fu_Zhang} and \onlinecite{Unitary_Matrix}
addressed the first issue, where the effect of the magnetic field
is absent. The Ref. \onlinecite{Fu_Zhang} uses the variational
method to study the Kondo effect, which is an approximation, thus
not a rigorous proof for the equivalence of the Kondo effect in
the topological insulator and graphene, although this study can
give fruitful information such as spin correlations. The Ref.
\onlinecite{Unitary_Matrix} also pointed out the equivalence based
on a different choice of the unitary transformation, and we
compare these two different transformations. However, the other
two issues have been discussed only in our study, and they are
essential to probe the surface states of topological insulators.

The present paper is organized as follows. In Sec. II we derive an
effective impurity action from the Anderson model, mapping the
model into an effective one dimensional action and integrating
over such one dimensional host electrons. Performing the
slave-boson mean-field analysis, we show that the peak splitting
of the impurity density of states under magnetic field differs
from the standard Zeeman energy proportional to the magnetic
field. In section III we present the host-electron LDOS and its
local charge density. They reveal that the decay of the Friedel
oscillation changes from an inverse-square law to an inverse law,
applying magnetic field. In section IV summary and discussion are
presented.

\section{Impurity dynamics}

\subsection{Equivalence between the Kondo effect of the helical
metal and the soft-gap Anderson model}

\subsubsection{An effective impurity action in the helical metal}

We start from an Anderson model in two dimensions
\begin{eqnarray}
H_{\text{hel}} &=&
\sum\limits_{\mathbf{k}\sigma\sigma^{\prime}}c_{\mathbf{k}\sigma
}^{\dagger} [
v_{F}(\mathbf{k}\cdot{\boldsymbol{\sigma}})_{\sigma\sigma^{\prime}}
-\mu_{\sigma} \delta_{\sigma\sigma^{\prime}} ]
c_{\mathbf{k}\sigma^{\prime}} +
{\displaystyle\sum\limits_{\sigma}} E_{f\sigma}
f_{\sigma}^{\dagger}f_{\sigma}
\nonumber \\
&+& {\displaystyle\sum\limits_{\mathbf{k}\sigma}} V\left(
\mathbf{k}\right) f_{\sigma}^{\dagger}c_{\mathbf{k}\sigma } +
\text{H.c.} + Un_{\uparrow}^{f}n_{\downarrow}^{f} . \label{helham}
\end{eqnarray}
$c_{\mathbf{k}\sigma }^{\dagger}$ ($c_{\mathbf{k}\sigma}$) is the
creation (annihilation) operator for host electrons with momentum
$\mathbf{k} = (k_{x}, k_{y})$. This metallic host is given by the
surface of the topological insulator, where the
$\mathbf{k}\cdot{\boldsymbol{\sigma}}$ term locks the electron
spin to its momentum, named as helical metal. $v_{\text{F}}$ is
the Fermi velocity, and $\sigma^{x}$, $\sigma^{y}$ are the Pauli
matrices. $\sigma=\pm 1$ is the spin index, and
$\mu_{\sigma}=\mu+\sigma h$ is an effective chemical potential,
where $h$ is the magnetic field. $f_{\sigma}^{\dagger}$
($f_{\sigma}$) represents the creation (annihilation) operator for
the magnetic impurity. $E_{f\sigma}=E_f-\sigma h$ is an effective
impurity energy level with the Zeeman energy contribution. $U$ is
the Coulomb interaction at the impurity site. $V(\mathbf{k})$ is
the hybridization between the impurity and helical electrons,
assumed to be $V(\mathbf{k})=V$ for simplicity.

In order to map this Hamiltonian to one dimensional model, we
write $\mathbf{k}=k( \cos(\phi), \sin(\phi))$, where
$k=|\mathbf{k}|$ and $\phi$ is the angle of the momentum
$\mathbf{k}$ from the $x$-axis. The Hamiltonian for helical
electrons without magnetic field can be rewritten as
\begin{equation}
H^{\text{c}}_{\text{hel}} =
\sum\limits_{\mathbf{k}\sigma\sigma^{\prime}} \varepsilon_{k}
c_{\mathbf{k}\sigma}^{\dagger} M_{\sigma\sigma^{\prime}} (\phi)
c_{\mathbf{k}\sigma^{\prime}}, \label{helhamc}
\end{equation}
where $\varepsilon_{k}=v_{\text{F}} k$ is the energy dispersion,
and
\begin{eqnarray}
\hat{M}\left( \phi\right)  =\left(
\begin{array}[c]{cc}%
0 & e^{-i\phi}\\
e^{i\phi} & 0
\end{array}
\right)
\end{eqnarray}
expresses the spin-momentum coupling. Introducing the unitary
transformation
\begin{eqnarray}
\hat{U}(\phi) =\frac{1}{\sqrt{2}}\left(
\begin{array}[c]{cc}
1 & 1\\
e^{i\phi} & -e^{i\phi}%
\end{array}
\right) ,
\end{eqnarray} the Hamiltonian (\ref{helhamc}) is diagonalized as
follows
\begin{equation}
H^{\text{c}}_{\text{hel}}= \sum\limits_{\mathbf{k}\sigma}
\varepsilon_{k}\sigma \gamma_{\mathbf{k}\sigma}^{\dagger
}\gamma_{\mathbf{k}\sigma}, \label{helhamc1}
\end{equation}
where
$\hat{\gamma}_{\mathbf{k}}=\hat{U}^{\dagger}(\phi)\hat{c}_{\mathbf{k}}$.
Expanding the $\gamma_{\mathbf{k}\sigma}$ operator in the basis of
angular-momentum eigenmodes \cite{Cassanello}
\begin{eqnarray}
\gamma_{\mathbf{k}\sigma}=\frac{1}{\sqrt{k}}\frac{1}{\sqrt{2\pi}}
\sum\limits_{m}e^{i m\phi}\gamma_{m\sigma}(k) ,
\end{eqnarray}
where $m$ is an integer representing angular momentum, we rewrite
the Hamiltonian (\ref{helhamc1}) as
\begin{eqnarray}
H^{\text{c}}_{\text{hel}}=
\sum\limits_{m\sigma}\int\limits_{0}^{\infty} d k
\varepsilon_{k}\sigma \gamma_{m\sigma}^{\dagger
}(k)\gamma_{m\sigma}(k) .
\end{eqnarray}

The hybridization term in Hamiltonian (\ref{helham}) becomes
\begin{eqnarray}
H^{\text{hyb}}_{\text{hel}}&=&\frac{\sqrt{N}V}{2\sqrt{\pi}}
\int\limits_{0}^{\infty} d k\sqrt{k}[
f_{\uparrow}^{\dagger}\gamma_{0\uparrow}(k)+f_{\uparrow
}^{\dagger}\gamma_{0\downarrow}(k) \nonumber \\
&+& f_{\downarrow}^{\dagger}\gamma_{-1\uparrow
}(k)-f_{\downarrow}^{\dagger}\gamma_{-1\downarrow}(k) ]
+\text{H.c.}
\end{eqnarray}
in the $\gamma_{m\sigma}$ basis, where $N$ is the number of
surface states. Note that only the $s$- ($m=0$) and $p$- ($m=-1$)
wave components contribute to the hybridization between the
impurity and conduction electrons. This is typical in the Dirac
spectrum,\cite{Cassanello} resulting from ($1$, $e^{i\phi}$) in
the unitary matrix while only the $s$-wave scattering is relevant
in the presence of the Fermi surface.

The Zeeman term is
\begin{eqnarray}
H^{h}_{\text{hel}} = -h \sum\limits_m \int\limits_{0}^{\infty} d k
[\gamma_{m\uparrow}^{\dagger}(k) \gamma_{m\downarrow}(k)+
\gamma_{m\downarrow}^{\dagger}(k) \gamma_{m\uparrow}(k) ]
\label{hham}
\end{eqnarray}
for helical electrons. Remember that the magnetic field is applied
in the $z$-direction. Since spins are locked along the momentum
direction in the plane, the magnetic field should flip the spin in
order to get its $z$-component. As a result, helical electrons at
the momentum $k_{F}$ couple to those at $-k_{F}$ in the presence
of magnetic field. As will be discussed in the next section, this
is the reason why the inverse square behavior in the Friedel
oscillation pattern turns into the inverse form. On the other
hand, this spin-flip process is not introduced into the
conventional soft-gap Anderson model by the magnetic field,
showing the inverse square law as long as the inter-valley
scattering is not taken into account.\cite{Bena}

It is convenient to consider
\begin{eqnarray}
\psi_{k\uparrow} &=& \frac{1}{\sqrt{2}} ( \gamma_{0\uparrow}(k)+
\gamma _{0\downarrow}(k) ), \\
\psi_{k\downarrow}&=& \frac{1}{\sqrt{2}} (
\gamma_{-1\uparrow}(k) - \gamma _{-1\downarrow}(k) ), \\
\varphi_{k\uparrow}&=& \frac{1}{\sqrt{2}}( \gamma_{0\uparrow}(k)
-\gamma _{0\downarrow}(k) ) ,\\
\varphi_{k\downarrow} &=& \frac{1}{\sqrt{2}} (
\gamma_{-1\uparrow}(k) + \gamma _{-1\downarrow}(k) ) .
\end{eqnarray}
One can verify that these operators satisfy the fermionic
anticommutation relations. Rewriting the Hamiltonian
(\ref{helham}) in this new basis, we obtain
\begin{eqnarray}
H_{\text{hel}}  &=& \sum\limits_{\sigma} \int\limits_{0}^{\infty}
d k \varepsilon_{k}
(\psi_{k\sigma}^{\dagger}\varphi_{k\sigma}+\varphi_{k\sigma
}^{\dagger}\psi_{k\sigma}) \nonumber \\
&-& \sum\limits_{\sigma} \int\limits_{0}^{\infty} d k
(\mu_{\sigma}\psi_{k\sigma}^{\dagger}\psi_{k\sigma}+\mu_{-\sigma
}\varphi_{k\sigma}^{\dagger}\varphi_{k\sigma}) \nonumber \\
&+& \sum\limits_{\sigma}E_{f\sigma}f_{\sigma}^{\dagger}f_{\sigma}
+U n_{\uparrow}^{f}n_{\downarrow}^{f} \nonumber
\\
&+& \sum\limits_{\sigma}\frac{\sqrt{N}V} {\sqrt{2\pi}}
\int\limits_{0}^{\infty} d
k\sqrt{k}f_{\sigma}^{\dagger}\psi_{k\sigma}+\text{H.c.} ,
\label{helhamf}
\end{eqnarray}
where the impurity couples to only $\psi_{\sigma}$-electrons.

Integrating over the $\varphi_{\sigma}$ fields, we obtain the
effective action
\begin{eqnarray}
S_{\text{hel}}  =\!\!
\sum\limits_{\sigma}\!\!\int\limits_{0}^{\beta}\!\!d\tau\!\!
\int\limits_{0}^{\beta}\!\!d\tau^{\prime}\!\!
\!\!\int\limits_{0}^{\infty}\!\! d k
\psi_{k\sigma}^{\dagger}(\tau)[g_{c\sigma}^{\text{hel}}
(k,\tau-\tau^{\prime})]^{-1}\psi_{k\sigma}(\tau^{\prime})\nonumber\\
 +\sum\limits_{\sigma}\int\limits_{0}^{\beta}d\tau
f_{\sigma}^{\dagger}
(\tau)[\partial_{\tau}+E_{f\sigma}]f_{\sigma}(\tau)+U\int\limits_{0}^{\beta
}d\tau n_{\uparrow}^{f}(\tau)n_{\downarrow}^{f}(\tau)\nonumber\\
+\sum\limits_{\sigma}\frac{\sqrt{N}V}{\sqrt{2\pi}}\int\limits_{0}^{\beta
}d\tau \int\limits_{0}^{\infty} d
k\sqrt{k}f_{\sigma}^{\dagger}(\tau)\psi_{k\sigma}(\tau)+\text{H.c.},
\;\;\;\;\;\;\;
\end{eqnarray}
where
\begin{eqnarray}
g_{c\sigma}^{\text{hel}}
(k,i\omega)=\frac{i\omega+\mu_{-\sigma}}{( i\omega +\mu_{\sigma})
(  i\omega+\mu_{-\sigma}) -(\varepsilon _{k})^{2}} \label{gc0}
\end{eqnarray}
is the $\psi_{\sigma}$ Green's function. Integration of the
$\psi_{\sigma}$ fields gives rise to an effective action for
dynamics of the magnetic impurity on the surface of the
topological insulator
\begin{eqnarray}
S_{\text{hel}} &=&\sum\limits_{\sigma}\int\limits_{0}^{\beta}d\tau f_{\sigma}^{\dagger}%
(\tau)[\delta(\tau-\tau^{\prime})(\partial_{\tau}+E_{f\sigma})\nonumber
\\
&+& \Delta_{\sigma
}^{\text{hel}}(\tau-\tau^{\prime})]f_{\sigma}(\tau^{\prime})
+U\int\limits_{0}^{\beta}d\tau
n_{\uparrow}^{f}(\tau)n_{\downarrow}^{f}(\tau) , \;\;\;
\label{helact}
\end{eqnarray}
where
\begin{eqnarray}
\Delta_{\sigma}^{\text{hel}}(i\omega)=\frac{N|V|^{2}}{2\pi}
\int\limits_{0}^{\infty} d k k g_{c\sigma}^{\text{hel}}(k,i\omega)
\end{eqnarray}
is the hybridization function. Using the Lorentzian
cutoff,\cite{Cassanello} we obtain the hybridization function
\begin{eqnarray}
\Delta_{\sigma}^{\text{hel}}(i\omega)=\frac{N|V|^{2}}{2\pi}
\int\limits_{0}^{\infty} d k k g_{c\sigma}^{\text{hel}}(k,i\omega)
\frac{2}{\pi}\frac{\Lambda v_{F}}{v_{F}^{2} k^2 + \Lambda^2} \nonumber \\
=-\Gamma \frac{2\Lambda(i\omega+\mu-\sigma h)}{(i\omega
+\mu)^{2}-h^{2}+\Lambda^{2}}\log\frac{\Lambda^{2}}{h^{2}-(i\omega+\mu)^{2}}
,
\end{eqnarray}
where $\Lambda$ is a cutoff in energy and $\Gamma=N|V|^2/4\pi^2
v_{F}$.

\subsubsection{An effective impurity action in the soft gap metal}

We consider the soft-gap Anderson model
\begin{eqnarray}
H_{\text{sg}} &=&
\sum\limits_{\mathbf{k}\alpha\alpha^{\prime}\sigma}c_{\mathbf{k}\alpha\sigma
}^{\dagger} [
v_{F}(\mathbf{k}\cdot{\boldsymbol{\sigma}})_{\alpha\alpha^{\prime}}
-\mu_{\sigma} \delta_{\alpha\alpha^{\prime}} ]
c_{\mathbf{k}\alpha^{\prime}\sigma} \nonumber \\
&+& {\displaystyle\sum\limits_{\sigma}} E_{f\sigma}
f_{\sigma}^{\dagger}f_{\sigma} +U
n_{\uparrow}^{f}n_{\downarrow}^{f}
\nonumber \\
&+& {\displaystyle\sum\limits_{\mathbf{k}\alpha\sigma}} V\left(
\mathbf{k}\right) f_{\sigma}^{\dagger}c_{\mathbf{k}\alpha\sigma
}+\text{H.c.}   . \label{sgham}
\end{eqnarray}
The notation in the Hamiltonian (\ref{sgham}) is the same as that
in Eq. (\ref{helham}) except for the conduction electrons which
now have an additional branch notation $\alpha, \alpha' =\pm 1$,
associated with the Pauli matrix
$\boldsymbol{\sigma}_{\alpha\alpha'}$. This may be interpreted as
pseudospin, which can result from two sublattices in graphene.
\cite{Neto} In addition to this pseudospin index, there exists
another Dirac cone in graphene, called valley. In this study we
take into account only one valley to focus on the comparison with
the helical metal. In the soft-gap metal momentum is not tied to
the electron spin, instead the pseudospin. This feature differs
from helical electrons.

Proceeding in the same way as in the previous subsection, we
obtain
\begin{eqnarray}
H_{\text{sg}}  &=& \sum\limits_{m=0,-1}\sum\limits_{\alpha\sigma}
\int\limits_{0}^{\infty} d k  k (\alpha
\varepsilon_{k}+\mu_{\sigma} )
\gamma_{m\alpha\sigma}^{\dagger}(k)\gamma_{m\alpha\sigma}(k)
\nonumber \\
&+& \sum\limits_{\sigma} E_{f\sigma}
f_{\sigma}^{\dagger}f_{\sigma}+
U n_{\uparrow}^{f}n_{\downarrow}^{f}\nonumber\\
&+& \frac{\sqrt{N}
V}{2\sqrt{\pi}}\sum\limits_{\sigma}\int\limits_{0}^{\infty} d k
\sqrt{k} f_{\sigma}^{\dagger}[\gamma_{0,+,\sigma}(k)+
\gamma_{0,-,\sigma}(k) \nonumber \\
&+& \gamma_{-1,+,\sigma}(k)-\gamma_{-1,-,\sigma}(k)]+\text{H.c.},
\label{sgham1}
\end{eqnarray}
where the $\gamma_{m\alpha\sigma}(k)$ basis diagonalizes the
conduction electron part of Hamiltonian (\ref{sgham}). Note that
the pseudospin $\alpha$ plays the same role as the spin of the
helical metal, where the pseudospin mixing appears in the
hybridization term.

We introduce $\psi_{1,\sigma}(k)=\gamma_{0,+,\sigma}(k)$ for $k
\geq 0$, and $\psi_{1,\sigma}(k)=\gamma_{0,-,\sigma}(k)$ for
$k<0$; $\psi_{2,\sigma}(k)=\gamma_{-1,+,\sigma}(k)$ for $k \geq
0$, and $\psi_{2,\sigma}(k)=-\gamma_{-1,-,\sigma}(k)$ for $k<0$.
These new operators allow us to rewrite the Hamiltonian
(\ref{sgham1}) in the following way
\begin{eqnarray}
H_{\text{sg}}  &=& \sum\limits_{\alpha=1,2;\sigma}%
\int\limits_{-\infty}^{\infty} d k
(\varepsilon_{k}-\mu_{\sigma})\psi_{\alpha\sigma}^{\dagger}(k)\psi
_{\alpha\sigma}(k) \nonumber
\\
&+& \sum\limits_{\sigma}E_{f\sigma}
f_{\sigma}^{\dagger}f_{\sigma}+Un_{\uparrow}^{f}n_{\downarrow}^{f} \nonumber\\
&+& \frac{\sqrt{N}V}{2\sqrt{\pi}}\sum\limits_{\alpha=1,2;\sigma}
\int\limits_{-\infty}^{\infty} d k
\sqrt{|k|}f_{\sigma}^{\dagger}\psi_{\alpha\sigma}(k)+\text{H.c.}
\;\;\;
\end{eqnarray}
Note that the magnetic field in the soft-gap metal couples to
electrons with the same spin and pseudospin. There is no spin-flip
or pseudospin-flip process, induced by magnetic field.

Integrating over the $\psi_{\alpha\sigma}$ field, we obtain an
effective impurity action from the soft-gap Anderson model
\begin{eqnarray}
S_{\text{sg}} &=& \sum\limits_{\sigma}\int\limits_{0}^{\beta}d\tau
f_{\sigma}^{\dagger}
(\tau)[\delta(\tau-\tau^{\prime})(\partial_{\tau}+E_{f\sigma})
\nonumber \\
&+& \Delta_{\sigma}^{\text{sg}}(\tau-\tau^{\prime})]
f_{\sigma}(\tau^{\prime})+U\int\limits_{0}^{\beta}d\tau
n_{\uparrow}^{f}(\tau)n_{\downarrow}^{f}(\tau), \;\;\;
\end{eqnarray}
where
\begin{eqnarray}
\Delta_{\sigma}^{\text{sg}}(i\omega) &=&
\frac{N|V|^{2}}{2\pi}\sum\limits_{\alpha}
\int\limits_{-\infty}^{\infty} dk |k|
\frac{1}{i\omega+\mu_{\sigma}-\varepsilon_{k}} \nonumber \\
&=& -\Gamma \frac{4\Lambda(i\omega+\mu_{\sigma})}{(i\omega
+\mu_{\sigma})^{2}+\Lambda^{2}}\log\frac{-\Lambda^{2}}{(i\omega+\mu_{\sigma
})^{2}}   \;\;\; \label{sgact}
\end{eqnarray}
is the hybridization function.

\subsubsection{Discussion}

It is quite interesting to see that the effective impurity action
in the helical metal is essentially identical with that in the
graphene case without magnetic field. The only difference is the
factor $2$ in the hybridization function of the soft-gap metal,
resulting from the pseudospin symmetry. This completes our proof
that the magnetic impurity on the surface of the topological
insulator is screened exactly, the same as the Kondo effect in the
soft-gap Anderson model.\cite{Cassanello} The identity of the
effective impurity action in the helical metal and graphene was
also pointed out in Ref. \onlinecite{Unitary_Matrix}.

An interesting point is the choice of the unitary matrix for
diagonalization. In this paper we choose the single-valued
function as the unitary matrix, given by $e^{i\phi}$. On the other
hand, the previous studies \cite{Fu_Zhang,Unitary_Matrix} took the
double-valued function $e^{i\phi/2}$ in the unitary matrix.
Although this kind of non-single-valued function can be utilized
in principle, the branch-cut should be taken into account with an
additional phase factor carefully. However, an effective impurity
action based on this double-valued unitary transformation is
completely identical with that based on the single-valued unitary
transformation. In appendix we perform the same procedure to find
the effective impurity action.

\subsection{U(1) slave-boson mean-field analysis in the presence of magnetic field}

Although the Kondo effect in the helical metal is completely the
same as that of the soft-gap Anderson model without magnetic
field, it becomes much different, applying magnetic field. This
originates from the fact that the magnetic field gives rise to the
spin-flip process in the helical metal while there is no such a
term in the graphene. This is reflected on the magnetic-field
dependence of the hybridization function. To reveal the effect of
the magnetic field, we use the slave-boson mean-field
approximation in the $U \rightarrow \infty$ limit. The electron
operator at the impurity site is decomposed as $f_{\sigma}=
d_{\sigma} \hat{b}^{\dagger}$, where $d_{\sigma}$ is a fermion
operator and $b$ is a boson one.\cite{ReadNewns} A constraint
should be imposed to match the Hilbert space of the original
electron representation with that of the slave-boson expression,
given by
\begin{eqnarray}
\hat{b}^{\dagger} \hat{b} + \sum_{\sigma} d_{\sigma}^{\dagger}
d_{\sigma} =1 . \label{cons}
\end{eqnarray}

\begin{figure}[t]
\includegraphics[width=0.4\textwidth]{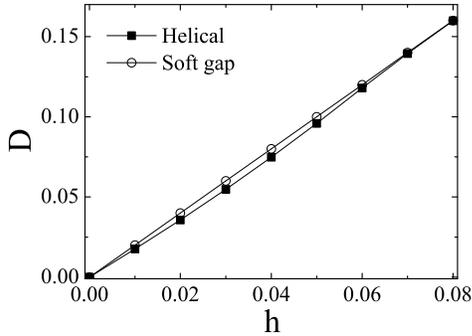}
\caption{The spin $\uparrow$ and $\downarrow$ splitting energy of
the Kondo peak under magnetic field in the spectral density of the
impurity : $\Gamma=1$, $\Lambda=50$, $\mu=0.1$, $T=0.01$,
$E_f=-6.2$ for the helical case and $E_f=-12.4$ for the soft gap
case.} \label{fig1}
\end{figure}

In the mean-field approximation the boson operator is replaced by
its average value $\langle \hat{b} \rangle = b$, and the
constraint equation (\ref{cons}) is taken into account by its
average version. The impurity Green's function is given by
\begin{eqnarray}
G_{d\sigma}(\omega) = \frac{1}{\omega - \tilde{E}_{f\sigma} - b^2
\Delta_{\sigma}(\omega)} ,
\end{eqnarray}
where $\tilde{E}_{f\sigma}=E_{f\sigma}+\lambda$ is an effective
energy level and $\lambda$ is the Lagrange multiplier to impose
the slave-boson constraint. We consider the chemical potential of
conduction electrons slightly above the Dirac point, i.e., $\mu >
0$. Fig. \ref{fig1} shows the splitting energy of the spin
$\uparrow$ and $\downarrow$ Kondo peaks in the spectral density of
the impurity under magnetic field. In the case of the soft-gap
Anderson model $D = 2 h$ results, as expected. On the other hand,
the Kondo-peak splitting in the helical metal deviates from the
typical $h$-linear behavior. This property distinguishes the Kondo
effect of the helical metal from that of the soft-gap Anderson
model and would be observed experimentally.

Physics behind this deviation lies in the spin locking along the
momentum direction in the plane. The energy cost induced by the
magnetic field in the $z$-direction becomes
\bqa && \Delta
E[\theta] = \alpha k_{F} S (1 - \cos \theta) - h S \sin \theta ,
\eqa
where the first contribution results from the spin-locking
kinetic term $\mathbf{k}\cdot{\boldsymbol{\sigma}}$ and the second
is the Zeeman energy. $S$ is the size of spin and $k_{F}$ is the
Fermi momentum. $\alpha$ is a positive numerical constant,
depending on the microscopic detail associated with the Fermi
surface geometry. $\theta$ is an angle from the $xy$ plane to the
$z$-direction, used to be the variational parameter. Minimizing
the energy with respect to $\theta$, we obtain
\bqa \cos \theta =
\frac{\alpha k_{F}}{\sqrt{h^{2} + (\alpha k_{F})^{2}}} , ~~~ \sin
\theta = \frac{h}{\sqrt{h^{2} + (\alpha k_{F})^{2}}} .
\eqa
As a
result, the peak splitting is given by \bqa &&  D(h) = 2 S
\Bigl(\sqrt{h^{2} + (\alpha k_{F})^{2}} - \alpha k_{F} \Bigr) ,
\eqa differentiated from the typical $D(h) = 2 S h$ behavior.

If the magnetic field lies in the $x$-direction, the energy cost
becomes
\begin{eqnarray}
\Delta E[\theta_\phi] =  \int\limits_{0}^{2\pi} \frac{d\phi}{2\pi}
[k_{F} S (\cos(\phi-\theta_\phi)-1) - h S \cos(\theta_\phi) ],
\end{eqnarray}
where $\theta_\phi$ is the angle between the spin and
$x$-direction. Minimizing the energy with respect to
$\theta_\phi$, we obtain
\begin{eqnarray}
\tan(\theta_\phi) =\frac{k_F \sin(\phi)}{k_F\cos(\phi)-h} .
\end{eqnarray}
The corresponding minimum energy becomes
\begin{eqnarray}
\Delta E= S \int\limits_{0}^{2\pi} \frac{d\phi}{2\pi}
\sqrt{k_F^2+h^2-2 k_F h \cos(\phi)} - S k_F.
\end{eqnarray}
Expanding the minimum energy in $h$, one can show that  the linear
term vanishes. As a result, the peak splitting $D(h) \sim h^2$ as
in the case of the magnetic field in the $z$-direction.

\section{Friedel oscillation in local charge and spin densities of conduction electrons}

In the previous section impurity dynamics was investigated,
considering two kinds of metallic hosts. We study its feedback
effect on conduction electrons around the magnetic impurity.

\subsection{Observable}

The presence of impurities spoils homogeneity of conduction
electrons, and their dynamics maintains local properties around
the impurity. Then, the Green's function of conduction electrons
is given by
\begin{eqnarray}
G_{c\sigma\sigma^{\prime}}(\mathbf{k},\mathbf{k}^{\prime},\tau) =
 - \langle T_{\tau} [c_{\mathbf{k}\sigma}(\tau)
 c^{\dagger}_{\mathbf{k}^{\prime}\sigma^{\prime}}(0)] \rangle ,
\end{eqnarray}
where two momentum indices appear.

LDOS of conduction electrons can be calculated via the Green's
function as follows
\begin{eqnarray}
\rho_{c\sigma}(\mathbf{r},\omega)&=&\sum\limits_{\mathbf{kk}^{\prime}
}e^{i(\mathbf{k}-\mathbf{k}^{\prime})\cdot\mathbf{r}} i
[G_{c\sigma\sigma}(\mathbf{k},\mathbf{k}^{\prime},\omega^{+})
\nonumber \\
&-& G_{c\sigma\sigma}(\mathbf{k},\mathbf{k}^{\prime},\omega^{-})]
, \label{ldos}
\end{eqnarray}
where $\omega^{\pm}=\omega\pm i 0^{+}$. The local charge density
can be also expressed via the Green's function
\begin{eqnarray}
n_{c\sigma}(\mathbf{r}) = T \sum\limits_{n}
\sum\limits_{\mathbf{kk}^{\prime}
}e^{i(\mathbf{k}-\mathbf{k}^{\prime})\cdot\mathbf{r}}
G_{c\sigma\sigma}(\mathbf{k},\mathbf{k}^{\prime},i\omega_{n})
e^{i\omega_n 0^{+}} , \label{chargedens}
\end{eqnarray}
where $\omega_n=(2n+1)\pi T$ is the Matsubara frequency.

The local spin density of states (LSDOS) is defined as
\begin{eqnarray}
\mathbf{s}_{c}(\mathbf{r},\omega)&=& \frac{1}{2}
\sum\limits_{\mathbf{kk}^{\prime}}
\sum\limits_{\sigma\sigma^{\prime}}
e^{i(\mathbf{k}-\mathbf{k}^{\prime})\cdot\mathbf{r}}i
\boldsymbol{\sigma}_{\sigma\sigma^{\prime}}[G_{c\sigma\sigma^{\prime}}
(\mathbf{k},\mathbf{k}^{\prime},\omega^{+})
\nonumber \\
&-&
G_{c\sigma\sigma^{\prime}}(\mathbf{k},\mathbf{k}^{\prime},\omega^{-})]
. \label{sdos}
\end{eqnarray}
The corresponding local spin density is also given by the Green's
function of conduction electrons
\begin{eqnarray}
\mathbf{S}_{c}(\mathbf{r}) \!=\! \frac{T}{2}\!\!
\sum\limits_{n\sigma\sigma^{\prime}}\!\!
\sum\limits_{\mathbf{kk}^{\prime}
}e^{i(\mathbf{k}-\mathbf{k}^{\prime})\cdot\mathbf{r}}
\boldsymbol{\sigma}_{\sigma\sigma^{\prime}}
G_{c\sigma\sigma^{\prime}}(\mathbf{k},\mathbf{k}^{\prime},i\omega_{n})
e^{i\omega_n 0^{+}} . \hspace{0.2cm}\label{sd}
\end{eqnarray}
In the following we evaluate all these quantities and find an
important fingerprint of the helical metal in the presence of
magnetic field.

\subsection{Friedel oscillation in the helical metal}

In the helical metal the conduction-electron Green's function can
be expressed in terms of the Green's functions of $\psi_{\sigma}$
and $\varphi_{\sigma}$ fields
\begin{eqnarray}
G_{c\sigma\sigma}^{\text{hel}}(\mathbf{k},\mathbf{k}^{\prime},\omega)
= \frac{1}{2\pi \sqrt{k k^{\prime}}} G_{\psi
\sigma}(k,k^{\prime},\omega) \nonumber \\
+ \frac{e^{-i(\phi-\phi^{\prime})\sigma}}{2\pi \sqrt{k
k^{\prime}}} G_{\varphi, -\sigma}(k,k^{\prime},\omega) \nonumber
\\
+  \sum\limits_{m\not=0,-1} \frac{\delta(k-k^{\prime})}{2\pi k}
g_{c\sigma}^{\text{hel}}(k,\omega) e^{-i m (\phi-\phi^{\prime})} ,
\label{gchel}
\end{eqnarray}
where $G_{\psi \sigma}(k,k^{\prime},\omega)$, $G_{\varphi,
\sigma}(k,k^{\prime},\omega)$ are the Green's functions of
$\psi_{\sigma}$ and $\varphi_{\sigma}$ fields, and
$g_{c\sigma}^{\text{hel}}(k,\omega)$ is given in Eq. (\ref{gc0}).
$\phi$ ($\phi^{\prime}$) is the angle of $\mathbf{k}$
($\mathbf{k}^{\prime}$) from the x-axis.

Based on the equation of motion method, the Green's functions of
$\psi_{\sigma}$ and $\varphi_{\sigma}$ fields can be expressed via
the impurity Green's function
\begin{eqnarray}
 G_{\psi\sigma}(k,k^{\prime},\omega) =
\delta(k-k^{\prime})g_{c\sigma}^{\text{hel}}(k,\omega) \nonumber \\
+ \frac{N |V|^2}{2\pi}\sqrt{kk^{\prime}
}g_{c\sigma}^{\text{hel}}(k,\omega)g_{c\sigma}^{\text{hel}}(k^{\prime}
,\omega)
G_{f\sigma}(\omega) ,  \label{psi}\\
 G_{\varphi\sigma}(k,k^{\prime},\omega)=
\frac{\delta(k-k^{\prime})}{\omega+\mu_{-\sigma}}
\nonumber \\
+\frac{v_{F}^{2}kk^{\prime}}{(\omega+\mu_{-\sigma})^{2}}
G_{\psi\sigma}(k,k^{\prime},\omega) , \label{phi}
\end{eqnarray}
where $G_{f\sigma}(\omega)$ is the impurity Green's function. Note
that Eqs. (\ref{psi})-(\ref{phi}) are exact. The slave-boson
mean-field approximation results in $G_{f\sigma}(\omega)= b^2
G_{d\sigma}(\omega)$.

Inserting the conduction-electron Green's function calculated by
Eqs. (\ref{gchel})-(\ref{phi}) into Eq. (\ref{ldos}), we obtain
the following expression
\begin{eqnarray}
\rho_{c\sigma}^{\text{hel}}(r,\omega) =
\rho_{0\sigma}^{\text{hel}}(\omega) + \Delta
\rho_{c\sigma}^{\text{hel}}(r,\omega) , \label{38}
\end{eqnarray}
where
\begin{eqnarray}
\rho_{0\sigma}^{\text{hel}}(\omega) = - \frac{1}{\pi} \text{Im}
\int\limits_{0}^{\infty} d k k g_{c\sigma}^{\text{hel}}(k,\omega^{+}) , \;\;\;\; \label{39}\\
\Delta\rho_{c\sigma}^{\text{hel}}(r,\omega) = -
\frac{\Gamma}{v_{F}} \text{Im} \Big[
  G_{f\sigma}(\omega^{+})
\Big(t_{0\sigma}^{\text{hel}}\big(r,\omega^{+}\big)
\Big)^{2} \nonumber \\
+G_{f,-\sigma}(\omega^{+})
\Big(t_{1}^{\text{hel}}\big(r,\omega^{+}\big)\Big)^{2}\Big]. \;\;
\label{40}
\end{eqnarray}
The functions of $t_{0\sigma}^{\text{hel}}(r,\omega)$ and
$t_{1}^{\text{hel}}(r,\omega)$  are defined as
\begin{eqnarray}
t_{0\sigma}^{\text{hel}}(r,\omega) &=&  v_{F}
\int\limits_{0}^{\infty} d k k J_{0}(k r)
\frac{1}{2}\sum_{s}[g^{\gamma}_{ss}(k,\omega) \nonumber
\\ &+& \sigma g^{\gamma}_{s,-s}(k,\omega)] , \label{41} \\
t_{1}^{\text{hel}}(r,\omega) &=&  v_{F} \int\limits_{0}^{\infty} d
k k J_{1}(k r) \frac{1}{2}\sum_{s}s g^{\gamma}_{ss}(k,\omega) ,
\label{42}
\end{eqnarray}
where $g^{\gamma}_{ss'}(k,\omega)$ is the bare Green's function of
the $\gamma_{\sigma}$ field
\begin{eqnarray}
\hat{g}^{\gamma}(k,\omega) = \left(
\begin{array}{cc}
\omega + \mu - \varepsilon_{k} & h \\
h & \omega + \mu + \varepsilon_{k}
\end{array} \right)^{-1} ,
\end{eqnarray}
and $J_n(x)$ is the Bessel function.\cite{mathbook}

The LDOS [Eq. (\ref{38})] of conduction electrons consists of two
contributions. $\rho_{0\sigma}^{\text{hel}}(\omega)$ [Eq.
(\ref{39})] corresponds to the homogeneous part, not involved with
impurity scattering. On the other hand,
$\Delta\rho_{c\sigma}^{\text{hel}}(r,\omega)$ [Eq. (\ref{40})]
results from impurity scattering, varying with the distance from
the impurity, which reflects modulation of excess charge around
the impurity. $t_{0\sigma}^{\text{hel}}(r,\omega)$ is associated
with the $s$-wave scattering channel ($m = 0$) while
$t_{1}^{\text{hel}}(r,\omega)$ is related with the $p$-wave ($m =
- 1$) channel, seen from $J_{0}(k r)$ and $J_{1}(k r)$,
respectively. The second contribution in
$t_{0\sigma}^{\text{hel}}(r,\omega)$ originates from spin mixing
induced by magnetic field.

Using the Lorentzian cutoff \cite{Cassanello} with $\Lambda$, we
obtain the $t$-functions
\begin{eqnarray}
t_{0\sigma}^{\text{hel}}(r,i\omega) = -\frac{2}{\pi}\frac{\Lambda
(i\omega+\mu-\sigma h)}{\Lambda^{2}-(\omega-i\mu)^{2}-h^{2}}
\nonumber \\
 \times \big[K_{0}\big(\frac{r}{v_{F}}\sqrt
{(\omega-i\mu)^{2}+h^{2}}\big)-K_{0}\big(\frac{r}{v_F}\Lambda\big)\big]
, \label{t0hel}\\
t_{1}^{\text{hel}}(r,i\omega)
=-\frac{2}{\pi}\frac{\Lambda}{\Lambda^{2}-(\omega-i\mu)^{2}-h^{2}}
\nonumber \\
\times \big[ \sqrt{(\omega-i\mu)^{2}+h^{2}}
K_{1}\big(\frac{r}{v_F}\sqrt{(\omega-i\mu)^{2}+h^{2}}\big) \nonumber \\
-\Lambda K_{1}\big(\frac{r}{v_F}\Lambda\big)\big] , \label{t1hel}
\end{eqnarray}
where $K_{n}(x)$ is the modified Bessel function. \cite{mathbook}
The modified Bessel function has the following asymptotic
expansion for its large argument \cite{mathbook}
\begin{eqnarray}
K_{n}(x) \sim \sqrt{\frac{\pi}{2 x}} e^{-x} \Big(1+ \frac{4
n^2-1}{8 x}\Big) . \label{asympt}
\end{eqnarray}

\begin{figure}[t]
\includegraphics[width=0.4\textwidth]{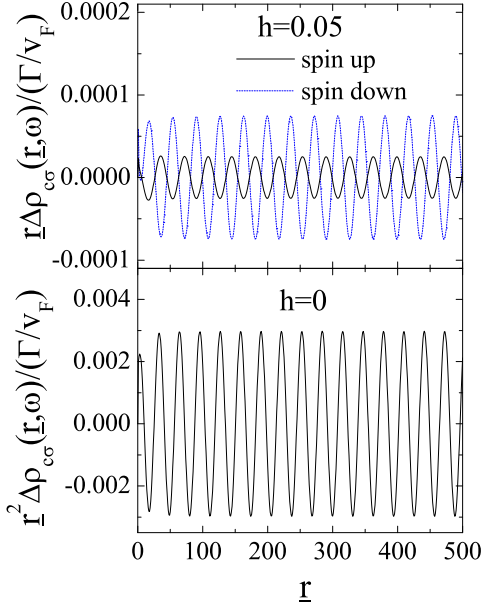}
\caption{(Color online) The LDOS of helical electrons $\Delta
\rho_{c\sigma}(\underline{r},\omega)$ with and without magnetic
field at $\omega=0$ ($\underline{r}=r/v_{F}$), scaled with
$1/\underline{r}^{\nu}$, where $\nu=1$ for finite magnetic field
$h$ and $\nu=2$ for $h=0$ ($\Gamma=1$, $\Lambda=50$, $\mu=0.1$,
$T=0.01$, $E_f=-6.2$). } \label{fig2}
\end{figure}

One can verify that the leading asymptotic behavior at large $r$
gives vanishing contributions to the LDOS in the absence of
magnetic field ($h=0$), providing $r |\omega+\mu|/v_{F} \gg 1$. In
other words, the $1/r$ contribution in
$t_{0\sigma}^{\text{hel}}(r,i\omega)$ is exactly cancelled by that
in $t_{1}^{\text{hel}}(r,i\omega)$. The next order of the
asymptotic expansion gives rise to the $1/r^2$ behavior with
oscillation, where its frequency is $2r(\omega+\mu)/v_{F}$, i.e.
\bqa && \Delta \rho_{c\sigma}^{\text{hel}}(r,\omega) \sim \sin(2 r
(\omega+\mu)/v_{F} )/r^2 . \nonumber \eqa When a weak magnetic
field is switched on, the leading asymptotic in Eq. (\ref{asympt})
maintains its non-vanishing contribution, corresponding to \bqa &&
\Delta \rho_{c\sigma}^{\text{hel}}(r,\omega) \sim \sin(2 r
(\omega+\mu)/v_{F} )/r . \nonumber \eqa There are three sources
which lead to the asymptotic $1/r$. The first is
$G_{f\sigma}(\omega) \not= G_{f,-\sigma}(\omega)$. The second is
the opening of an effective gap due to the magnetic field. The
last is the opposite spin scattering contribution, given by
$\sigma g^{\gamma}_{s,-s}(k,\omega)$ in Eq. (\ref{41}). Then, the
$1/r$ contribution in Eq. (\ref{41}) does not cancel that of Eq.
(\ref{42}).

In Fig. \ref{fig2} we plot $r^{\nu} \Delta
\rho_{c\sigma}^{\text{hel}} (r,\omega)$, where $\nu=2$ for $h=0$,
and $\nu=1$ for finite $h$. The impurity Green's function is
calculated within the slave-boson mean-field approximation,
described in the previous section. This plot confirms the above
analysis for the asymptotic behavior of the LDOS at large distance
$r$. The LDOS exhibits the Friedel oscillation, where its decay
changes from $1/r^2$ to $1/r$ when the magnetic field is switched
on. The magnetic field distinguishes the asymptotic behavior of
the LDOS unambiguously.

\begin{figure}[t]
\includegraphics[width=0.4\textwidth]{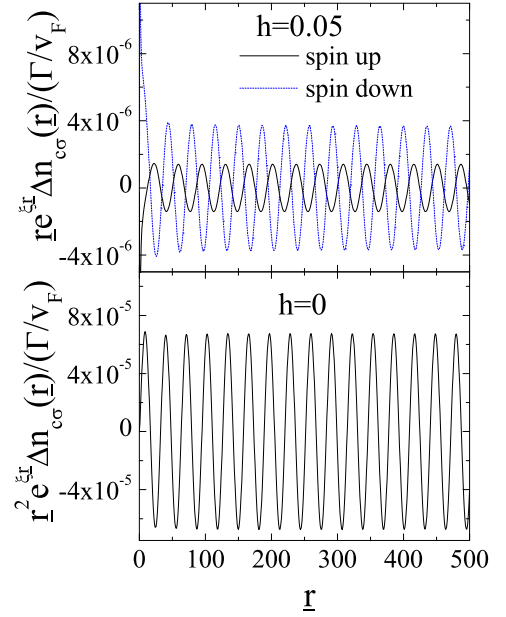}
\caption{(Color online) The local charge density of helical
electrons $\Delta n_{c\sigma}(\underline{r})$ with and without
magnetic field ($\underline{r}=r/v_{F}$), scaled with $\exp(-\xi
\underline{r})/\underline{r}^{\nu}$, where $\nu=1$, $\xi\approx
0.0712$ for $h=0.05$ and $\nu=2$, $\xi\approx 0.0628$ for $h=0$
($\Gamma=1$, $\Lambda=50$, $\mu=0.1$, $T=0.01$, $E_f=-6.2$). }
\label{figcdhel}
\end{figure}

The local charge density can be found from Eq. (\ref{chargedens})
\begin{eqnarray}
n_{c}^{\text{hel}}(r)=n_{0\sigma}^{\text{hel}} + \Delta
n_{c\sigma}^{\text{hel}}(r) ,
\end{eqnarray}
where $n_{0\sigma}^{\text{hel}}$ is the bare charge density of
helical electrons, and $\Delta n_{c\sigma}^{\text{hel}}(r)$ is the
local modulation of the helical-electron charge density due to the
presence of an impurity. $\Delta n_{c\sigma}^{\text{hel}}(r)$
corresponds to the $\Delta\rho_{c\sigma}^{\text{hel}}(r,\omega)$
part in the LDOS. At large distance $r$, we find that $\Delta
n_{c\sigma}^{\text{hel}}(r)$ fits very well with the asymptotic
expansion $\exp(-\xi r/v_{F})/r^{\nu}$, where $\nu$ is the same
power of the LDOS. $\xi$ is not a universal parameter, depending
on the microscopic detail. In Fig. \ref{figcdhel} we plot $r^{\nu}
\exp(\xi r/v_{F}) \Delta n_{c\sigma}^{\text{hel}}(r)$ as a
function of $r$. It shows that the local charge density also
exhibits the Friedel oscillation, where its decay, apart from the
relaxation part $\exp(-\xi r/v_{F})$, also obeys the same power
law as the LDOS.

In general, the LSDOS of helical electrons does not vanish. The
$z$-component of the LSDOS is basically the difference of the LDOS
for up and down spins. It vanishes when magnetic field is absent.
When the magnetic field is switched on, like the LDOS, the
$z$-component of the LSDOS also exhibits the Friedel oscillation
and decays as $1/r$ at large distance $r$. The local spin density
$S^{z}_{c}(r)$ also has the asymptotic $\exp(-\xi r/v_F)/r$ when
the magnetic field is finite. The $x$- and $y$- components of the
LSDOS are finite even in the absence of magnetic field as it
should be.

\subsection{Friedel oscillation in the soft-gap metal}

One can calculate the LDOS of conduction electrons in the soft-gap
metal, following the previous discussion in the helical metal. The
electron Green's function can be expressed in the
$\gamma_{\sigma}$ basis of Hamiltonian (\ref{sgham1})
\begin{eqnarray}
G_{c\alpha\beta\sigma}^{\text{sg}}(\mathbf{k},\mathbf{k}^{\prime},\omega)
\equiv \langle\langle c_{\mathbf{k}\beta\sigma}|
c^{\dagger}_{\mathbf{k}\alpha\sigma}\rangle\rangle_{\omega}
\nonumber \\
=\frac{\Phi_{\alpha\beta}(\phi,\phi^{\prime})}{4\pi\sqrt{kk^{\prime}}}
\sum\limits_{mm^{\prime}=0,-1}
\sum\limits_{\delta\tau}e^{im\phi-im^{\prime}\phi^{\prime}}
\eta_{\delta\tau}^{\alpha\beta} \nonumber \\
\langle\langle
\gamma_{m\delta\sigma}(k)|\gamma_{m^{\prime}\tau\sigma}^{\dagger}
(k^{\prime})\rangle\rangle_{\omega} \nonumber \\
+\frac{\Phi_{\alpha\beta}(\phi,\phi^{\prime})}{4\pi k}
\delta(k-k^{\prime}) \sum\limits_{mm^{\prime}\not=0,-1} e^{i m
(\phi-\phi^{\prime})} \nonumber \\
\Big(\frac{1}{\omega+\mu_{\sigma
}-v_{F}k}+\alpha\beta\frac{1}{\omega+\mu_{\sigma}+v_{F}k} \Big),
\label{sggreen}
\end{eqnarray}
where
$\Phi_{\alpha\beta}(\phi,\phi^{\prime})=\exp[i(1-\alpha)\phi/2-i(1-\beta)\phi^{\prime}/2]$
and $\eta_{++}^{\alpha\beta}=1$, $\eta_{+-}^{\alpha\beta}=\beta$,
$\eta _{-+}^{\alpha\beta}=\alpha$,
$\eta_{--}^{\alpha\beta}=\alpha\beta$.

Using the equation of motion method, the $\gamma_{\sigma}$ Green's
function is expressed by the impurity Green's function
\begin{eqnarray}
\langle\langle
\gamma_{m\delta\sigma}(k)|\gamma_{m^{\prime}\tau\sigma}^{\dagger
}(k^{\prime})\rangle\rangle _{\omega}
=\delta(k-k^{\prime})\delta_{mm^{\prime}}\delta_{\delta\tau}
g^{\gamma}_{\delta\sigma}(k,\omega) \nonumber \\
+\frac{N|V|^{2}}{4\pi}\sqrt{kk^{\prime}} \zeta_{m\delta}
\zeta_{m^{\prime}\tau} g^{\gamma}_{\delta\sigma}(k,\omega)
g^{\gamma}_{\tau\sigma}(k^{\prime},\omega) G_{f\sigma}(\omega) ,
\label{sggamma}
\end{eqnarray}
where $\zeta_{0+}=\zeta_{0-}=\zeta_{-1+}=-\zeta_{-1-}=1$ and
\begin{eqnarray}
g_{\alpha\sigma}^{\gamma}(k,\omega)
=\frac{1}{\omega+\mu_{\sigma}-\alpha \varepsilon_k} .
\end{eqnarray}

Inserting the conduction-electron Green's function calculated by
Eqs. (\ref{sggreen})-(\ref{sggamma}) into Eq. (\ref{ldos}), we
obtain the LDOS
\begin{eqnarray}
\rho_{c\sigma}^{\text{sg}}(\mathbf{r},\omega) &=& i \sum_{\alpha}
\sum_{\mathbf{k}\mathbf{k}^{\prime}}
e^{i(\mathbf{k}-\mathbf{k}^{\prime}) \cdot \mathbf{r}}
[G^{\text{sg}}_{c\alpha\alpha\sigma}(\mathbf{k},
\mathbf{k}^{\prime},\omega^{+}) \nonumber \\
&& \hspace{2.3cm} -G^{\text{sg}}_{c\alpha\alpha\sigma}(\mathbf{k},
\mathbf{k}^{\prime},\omega^{-})] \nonumber \\
&=&\rho_{0\sigma}^{\text{sg}}(\omega) + \Delta
\rho_{c\sigma}^{\text{sg}}(r,\omega) ,
\end{eqnarray}
where
\begin{eqnarray}
\rho_{0\sigma}^{\text{sg}}(\omega) = -\frac{1}{\pi} \text{Im}
\int\limits_{0}^{\infty} d k k g_{c\sigma}^{\text{sg}}(k,\omega^{+}) , \;\;\;\; \\
\Delta\rho_{c\sigma}^{\text{sg}}(r,\omega) = -
\frac{4\Gamma}{v_{F}} \text{Im} \Big[ G_{f\sigma}(\omega^{+})
\Big(\big(t_{0\sigma}^{\text{sg}}\big(r,\omega^{+}\big) \big)^{2}
  \nonumber \\
 + \big(t_{1\sigma}^{\text{sg}}\big(r,\omega^{+}\big)\big)^{2}  \Big) \Big]. \;\;
\end{eqnarray}
The functions $t_{0\sigma}^{\text{sg}}(r,\omega)$ and
$t_{1\sigma}^{\text{sg}}(r,\omega)$ are given by
\begin{eqnarray}
t_{0\sigma}^{\text{sg}}(r,\omega) &=&  v_{F}
\int\limits_{0}^{\infty} d k k J_{0}(k r)
\frac{1}{2}\sum_{\alpha} g^{\gamma}_{\alpha\sigma}(k,\omega) , \\
t_{1\sigma}^{\text{sg}}(r,\omega) &=&  v_{F}
\int\limits_{0}^{\infty} d k k J_{1}(k r)
\frac{1}{2}\sum_{\alpha}\alpha g^{\gamma}_{\alpha\sigma}(k,\omega)
,
\end{eqnarray}
and the bare Green's function of conduction electrons is
\begin{eqnarray}
g_{c\sigma}^{\text{sg}}(k,\omega) =\sum_{\alpha}
g^{\gamma}_{\alpha\sigma}(k,\omega) .
\end{eqnarray}
Note that the magnetic field in the $z$-direction does not cause
spin flipping in the graphene, thus the gap opening does not
occur, differentiated from the helical metal.

\begin{figure}[b]
\includegraphics[width=0.4\textwidth]{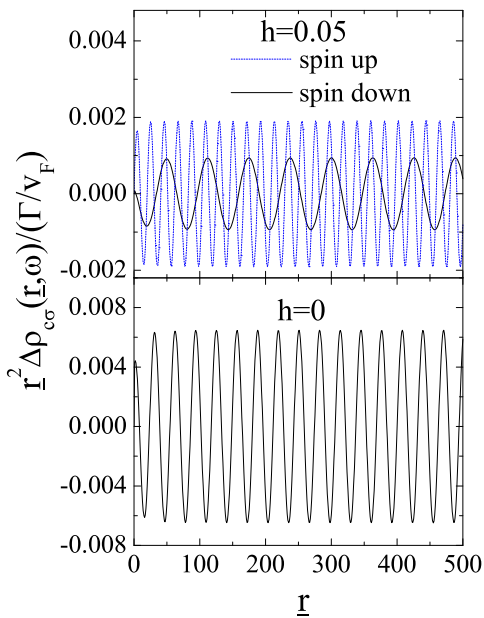}
\caption{(Color online) The LDOS of soft-gap electrons $\Delta
\rho_{c\sigma}(\underline{r},\omega)$ with and without magnetic
field at $\omega=0$ ($\underline{r}=r/v_{F}$), scaled with
$1/\underline{r}^{2}$ ($\Gamma=1$, $\Lambda=50$, $\mu=0.1$,
$T=0.01$, $E_f=-12.4$). } \label{fig3}
\end{figure}

Using the Lorentzian cutoff \cite{Cassanello} with $\Lambda$, we
obtain the $t$-functions
\begin{eqnarray}
t_{0\sigma}^{\text{sg}}(r,i\omega) =
-\frac{2}{\pi}\frac{\Lambda(i\omega+\mu_{\sigma})}{\Lambda^{2}-(\omega-i\mu_\sigma)^{2}}
\nonumber \\
\times
\big[K_{0}\big(\frac{r}{v_F}(\omega-i\mu_\sigma)\big)-
K_{0}\big(\frac{r}{v_F}\Lambda\big)\big], \label{t0sg}\\
t_{1\sigma}^{\text{sg}}(r,i\omega)
=-\frac{2}{\pi}\frac{\Lambda}{\Lambda^{2}-(\omega-i\mu_\sigma)^{2}}
\nonumber \\
\times \big[ (\omega-i\mu_\sigma)
K_{1}\big(\frac{r}{v_F}(\omega-i\mu_\sigma)\big) -\Lambda
K_{1}\big(\frac{r}{v_F}\Lambda\big)\big] . \label{t1sg}
\end{eqnarray}
One can verify that the leading order in the asymptotic expansion
of the modified Bessel function in Eq. (\ref{asympt}) gives
vanishing contributions to the LDOS for any values of $h$. The
next order of the asymptotic expansion leads to $\Delta
\rho_{c\sigma}^{\text{sg}} \sim
\sin(2r(\omega+\mu_{\sigma})/v_F)/r^2$. This situation is exactly
the same as the intranodal scattering in graphene.\cite{Bena} In
the present model we take into account only one Dirac cone, thus
internodal scattering is not introduced. It was demonstrated that
the $1/r$ behavior can originate from the internodal scattering
between two different valleys in graphene. The inter-valley
scattering gives rise to pseudospin mixing, resulting in the $1/r$
behavior. In the helical metal the magnetic field gives rise to
spin flipping, allowing the $1/r$ law.

In Fig. \ref{fig3} we plot $r^2 \Delta
\rho_{c\sigma}^{\text{sg}}(r,\omega)$, where the impurity Green's
function is based on the slave-boson mean-field analysis. This
plot confirms the $1/r^2$ decay of the LDOS, independent on the
magnetic field. In particular, we point out
$t_{n}^{\text{hel}}(r,\omega)=t_{n}^{\text{sg}}(r,\omega)$ in the
absence of magnetic field, explicitly verified from Eqs.
(\ref{t0hel})-(\ref{t1hel}) and (\ref{t0sg})-(\ref{t1sg}),
resulting in the completely same Friedel oscillation for both
helical and graphene cases.  One may regard that this equivalence
originates from the identical effective impurity action for both
helical and graphene cases without the magnetic field. On the
other hand, the magnetic field leads the oscillation frequency of
the $\downarrow$ spin to differ from that of the $\uparrow$ spin,
where the period difference is proportional to the magnetic field
strength, an important different point from the helical metal.

\begin{figure}[t]
\includegraphics[width=0.4\textwidth]{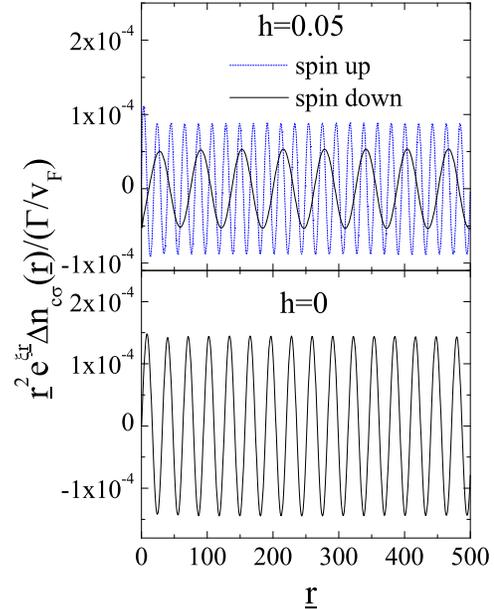}
\caption{(Color online) The local charge density of soft-gap
electrons $\Delta n_{c\sigma}(\underline{r})$ with and without
magnetic field ($\underline{r}=r/v_{F}$), scaled with $\exp(-\xi
\underline{r})/\underline{r}^{2}$, where $\xi\approx 0.0628$ (
$\Gamma=1$, $\Lambda=50$, $\mu=0.1$, $T=0.01$, $E_f=-12.4$). }
\label{figcdsoft}
\end{figure}

We also calculate the local charge density of soft-gap electrons.
In Fig. \ref{figcdsoft} we plot $r^{\nu} \exp(\xi r/v_{F}) \Delta
n_{c\sigma}^{\text{sg}}(r)$, where $\nu=2$, and $\xi$ is an
asymptotic fitting parameter. It confirms the asymptotic behavior
of the local charge density as $\exp(-\xi r/v_F)/r^{\nu}$. As the
LDOS, the decay of the local charge density always obeys the
inverse-square law independently on magnetic field, apart from the
relaxation part $\exp(-\xi r/v_F)$.

The $z$-component LSDOS also vanishes in the absence of magnetic
field as the helical metal. When the magnetic field is turned on,
the $z$-component of the LSDOS also decays as $1/r^2$ at large
distance $r$. Since the Friedel oscillations in the LDOS for up
and down spins have different periods, it should exhibit amplitude
modulation, not shown in the helical metal. One can show that
$S^{z}_{c}(r)$ obeys the asymptotic expression $\exp(-\xi
r/v_F)/r^2$ for finite magnetic fields while the other components
always vanish, different from the helical metal.

 One may ask why the Friedel oscillation reflects
spin or pseudospin physics although it is basically involved with
charge dynamics. The underlying mechanism is the coupling between
the orbital motion of charge degrees of freedom and spin or
pseudospin dynamics. To control spin dynamics with magnetic field
changes the orbital motion of charge, or to modify charge dynamics
by electric field governs the spin dynamics via the spin-orbit
coupling physics. This is one of the main research directions in
the present condensed matter physics.

\section{Discussion and Summary}

In this paper we investigated the Kondo effect and the associated
Friedel oscillation on the surface of the topological insulator,
where spins are locked along the momentum direction. In
particular, we examined the role of magnetic field $h$ to
distinguish the Kondo effect and Friedel oscillation of the
helical metal from those of the graphene metal, where pseudospins
are locked along the momentum direction. It turns out that both
the Kondo effect and Friedel oscillation of the helical metal are
completely identical to those of the standard soft-gap Anderson
model in the absence of magnetic field. However, the magnetic
field was shown to play a different role for each case.

We found that the spin $\uparrow$ and $\downarrow$ splitting
energy of the Kondo peak does not follow the typical $h$-linear
behavior of the soft-gap Anderson model. We revealed its physical
origin and derived the analytic expression. In addition, we showed
that the Friedel oscillation of the helical metal changes from the
typical $1/r^{2}$ law associated with the Dirac spectrum to the
$1/r$ behavior, applying magnetic field in the $z$-direction while
it still remains as $1/r^{2}$ in the graphene case. We clarified
physics behind this change that magnetic field gives rise to spin
mixing between $\uparrow$ and $\downarrow$ in the helical metal,
resulting in gap. We pointed out that this mechanism is quite
analogous to the internodal scattering in the graphene case,
\cite{Bena} introducing the pseudospin flipping. We propose the
Kondo effect and Friedel oscillation as the fingerprint for the
surface state of the topological insulator, measurable by Fourier
transformation scanning tunneling
spectroscopy.\cite{Mallet,Rutter,Vonau,Dupont}

 An interesting issue not discussed in this study is
on Ruderman-Kittel-Kasuya-Yosida (RKKY) correlations. It was
demonstrated based on the variational wave-function approach that
the spin-orbital quenching results in anisotropy for spin
dynamics, leading spin correlations of the topological surface
different from those of graphene which are SU(2)
symmetric.\cite{Fu_Zhang} This nontrivial spin dynamics in the
topological surface gives an interesting problem. Increasing
magnetic impurities, which kinds of spin dynamics will appear as a
competition between the Kondo effect and RKKY correlation? This
research direction opens a novel window for the interplay among
interactions (Kondo and RKKY), disorder, and
topology.\cite{Coleman,Tripathi,Tran_Kim_DisALM}

\begin{acknowledgments}

We would like to thank T. Takimoto for useful discussions. We
acknowledge the Korea Ministry of Education, Science and
Technology (MEST) for the grant of the National Research
Foundation of Korea (No. 2010-0074542) at the Asia Pacific Center
for Theoretical Physics. M.-T. was also supported by the National
Foundation for Science and Technology Development (NAFOSTED) of
Vietnam.

\end{acknowledgments}

\appendix*

\section{Comment on the unitary transformation}

We consider another choice of the unitary transformation.
\cite{Fu_Zhang,Unitary_Matrix} The helical electron Hamiltonian
(\ref{helhamc}) can be diagonalized by the unitary transformation
\begin{eqnarray}
\hat{U}(\phi)=\frac{1}{\sqrt{2}}\left(
\begin{array}[c]{cc}
e^{-i\phi/2} & e^{-i\phi/2}\\
e^{i\phi/2} & -e^{i\phi/2}%
\end{array}
\right) .
\end{eqnarray}
This unitary transformation matrix has its period $4\pi$, and the
Berry phase $\pi$ is acquired when an electron encircles the Fermi
surface. It has been chosen to study the Kondo effect in the
helical metal, based on the variational method. \cite{Fu_Zhang}

Proceeding in a similar way as in Sec. II, we obtain
\begin{eqnarray}
H^{c}_{\text{hel}}=\sum\limits_{m\sigma} \int\limits_{0}^{\infty}
d k \varepsilon_{k} \sigma \gamma_{m\sigma}^{\dagger}(k)
\gamma_{m\sigma}(k) .
\end{eqnarray}
The hybridization Hamiltonian becomes
\begin{eqnarray}
H_{\text{hel}}^{\text{hyb}}=\frac{\sqrt{N}V}{\sqrt{2\pi}}%
\sum\limits_{m \sigma s} \int\limits_{0}^{\infty} d k \sqrt{k}
f_{\sigma}^{\dagger} \xi_{\sigma s}(m) \gamma_{m s}(k)+
\text{H.c.} , \hspace{0.5cm}
\end{eqnarray}
where
\begin{eqnarray}
\hat{\xi}(m) = \left(
\begin{array}
[c]{cc}%
\xi_{m}^{-} & \xi_{m}^{-} \\
\xi_{m}^{+} & -\xi_{m}^{+}
\end{array}
\right)
\end{eqnarray}
and \bqa && \xi^{\pm}_{m}= \frac{\sqrt{2} }{\pi}\frac{i}{2 m \pm
1} . \nonumber \eqa It should be noted that the hybridization
Hamiltonian in this $\gamma_{\sigma}$ basis couples the impurity
for all angular momentum $m$, much different from the case of the
helical metal.

We introduce new operators for diagonalization
\begin{eqnarray}
\psi_{m\pm}(k)= \frac{1}{\sqrt{2}} (\gamma_{m\uparrow}(k) \pm
\gamma_{m\downarrow}(k) ) .
\end{eqnarray}
Then, the Hamiltonian (\ref{helham}) is written as follows
\begin{eqnarray}
H_{\text{hel}} &=& \sum\limits_{m} \int\limits_{0}^{\infty} d k
\varepsilon_{k}[\psi_{m+}^{\dagger}(k)\psi_{m-}(k)+
\psi_{m-}^{\dagger}(k)\psi_{m+}(k)] \nonumber \\
&-& \sum\limits_{m\sigma} \int\limits_{0}^{\infty} d k
\mu_{\sigma} \psi_{m\sigma}^{\dagger}(k) \psi_{m\sigma}(k)  \nonumber \\
&+& \sum\limits_{\sigma} E_{f\sigma}
f_{\sigma}^{\dagger}f_{\sigma} +
U n_{\uparrow}^{f}n_{\downarrow}^{f} \nonumber \\
&+& \frac{\sqrt{N} V}{\sqrt{\pi}} \sum\limits_{m\sigma}
\int\limits_{0}^{\infty} d k \sqrt{k}
f_{\sigma}^{\dagger}\xi_{m}^{-\sigma}\psi_{m\sigma}(k)+\text{H.c.}
\end{eqnarray}

Integrating over the $\psi_{\sigma}$ field, we obtain the
effective impurity action
\begin{eqnarray}
S_{\text{hel}} &=&\sum\limits_{\sigma\sigma^{\prime}}
\int\limits_{0}^{\beta}d\tau f_{\sigma}^{\dagger}%
(\tau)[\delta_{\sigma\sigma^{\prime}}
\delta(\tau-\tau^{\prime})(\partial_{\tau}+E_{f\sigma})\nonumber
\\
&+&\tilde{\Delta}_{\sigma\sigma^{\prime}}(\tau-\tau^{\prime})]
f_{\sigma^{\prime}}(\tau^{\prime}) +U\int\limits_{0}^{\beta}d\tau
n_{\uparrow}^{f}(\tau)n_{\downarrow}^{f}(\tau) , \hspace{0.8cm}
\label{helactappend}
\end{eqnarray}
where
\begin{eqnarray}
\tilde{\Delta}_{\sigma\sigma^{\prime}}(i\omega)=\frac{N|V|^{2}}{\pi}
\sum_{m} \int\limits_{0}^{\infty} d k k
\tilde{g}_{\sigma\sigma^{\prime}}(k,i\omega) \xi_{m}^{-\sigma}
(\xi_{m}^{-\sigma^{\prime}})^{*}, \hspace{0.5cm}
\end{eqnarray}
and
\begin{eqnarray}
\tilde{g}_{\sigma\sigma^{\prime}}(k,i\omega) = \left(
\begin{array}{cc}
i\omega + \mu + h & \varepsilon_{k} \\
\varepsilon_{k} & i\omega + \mu - h
\end{array} \right)^{-1}_{\sigma\sigma^{\prime}} .
\end{eqnarray}
Calling $\sum_{m} \xi_{m}^{-\sigma}
(\xi_{m}^{-\sigma^{\prime}})^{*} =\delta_{\sigma\sigma^{\prime}}/2
$, we obtain \bqa &&
\tilde{\Delta}_{\sigma\sigma^{\prime}}(i\omega) =
\delta_{\sigma\sigma^{\prime}}
\Delta_{\sigma}^{\text{hel}}(i\omega) . \eqa Thus, the effective
action [Eq. (\ref{helactappend})] is completely identical to the
one [Eq. (\ref{helact})].

The reason why the single-valuedness of the unitary transformation
is not relevant for impurity dynamics may be the fact that the
Berry phase contribution is integrated out, not affecting the
local dynamics. Then, the feedback effect of the impurity dynamics
to conduction electrons may be modified by the double-valued
unitary transformation. Actually, all angular-momentum channels
are coupled to the impurity. This will be addressed near future.

\end{document}